\documentclass[fleqn,usenatbib]{mnras}

\usepackage{newtxtext,newtxmath}
\usepackage{graphicx} 
\usepackage{cancel}

\usepackage[T1]{fontenc}

\DeclareRobustCommand{\VAN}[3]{#2}
\let\VANthebibliography\thebibliography
\def\thebibliography{\DeclareRobustCommand{\VAN}[3]{##3}\VANthebibliography}

\usepackage{graphicx}	
\usepackage{amsmath}	

\title[Photometric Lightcurve Intercalibration using Comparison Stars]{PyTICS: An Iterative Method for Photometric Lightcurve Intercalibration using Comparison Stars}

\author[R. Vieliute et al.]{
Roberta Vieliute$^{1}$, Juan V. Hernández Santisteban$^{1}$, Keith Horne$^{1}$ and Hannah Cornfield$^{1,2}$
\\
$^{1}$SUPA School of Physics and Astronomy, University of St Andrews, St Andrews, Scotland, KY16 9SS, UK \\
$^{2}$Department of Physics and Astronomy, University of Southampton, Highfield Campus, Southampton, SO17 1BJ, UK
}

\date{Accepted XXX. Received YYY; in original form ZZZ}

\pubyear{\the\year{}}

\begin{document}
\label{firstpage}
\pagerange{\pageref{firstpage}--\pageref{lastpage}}
\maketitle

\begin{abstract}
Intensive reverberation mapping monitoring programs combine ground-based photometric observations from different telescopes, requiring intercalibration of lightcurves to reduce systematic instrumental differences. We present a new iterative algorithm to calibrate photometric time-series data of active galactic nuclei (AGN) using 100s of comparison stars on the same images, building upon the established method of ensemble photometry. The algorithm determines telescope-specific and epoch-specific correction parameters, and simultaneously computes a multi-component noise model to account for underestimated uncertainties based on the scatter in the comparison star data, effectively identifying problematic epochs, telescopes, and stars. No assumptions need to be made about the AGN variability shape, and the algorithm can in principle be applied to any astronomical object. We demonstrate our method on lightcurves taken with ten 1-m telescopes from the Las Cumbres Observatory (LCO) robotic telescope network. Comparing our results to other intercalibration tools, we find that the algorithm can more accurately quantify the uncertainties in the data. We describe additional corrections that can be made for particularly bluer AGNs like Fairall\,9, arising due to systematic effects dependent on star colour.
\end{abstract}

\begin{keywords}
galaxies: active -- methods: statistical
-- techniques: photometric
\end{keywords}



\section{Introduction}

Emission from active galactic nuclei (AGN) is inherently highly variable and spans nearly the entire
electromagnetic spectrum, as both continua and emission lines. These time-dependent, multi-wavelength characteristics allow for the application of reverberation mapping (RM): a technique that uses time-series observations to probe the size and structure of AGN components that are otherwise too compact to resolve spatially with current telescopes (see Cackett et al. \citeyear{cackett_reverberation_2021} for a recent review). Time delays among correlated flux changes in different wavelength bands are interpreted as the light-crossing time between different emitting regions, giving a first-order estimate for their size $R$\,$\approx$\,$\tau c$, assuming a common impulse signal. The RM technique was first used to study the broad line region (BLR) clouds by measuring time delays between flux variations of the broad emission lines and the UV/optical continuum (e.g., 
Blandford \& McKee \citeyear{blandford_reverberation_1982}; 
Peterson \citeyear{peterson_reverberation_1993}; Shen et al.
\citeyear{shen_sloan_2024}). 
It has since been successfully extended to the other constituents, including the thermally emitting accretion disc using multi-wavelength UV/optical photometric observations (e.g., Krolik et al. \citeyear{krolik_ultraviolet_1991}; Fausnaugh et al. \citeyear{fausnaugh_continuum_2017}).

Even though photometric interband delays have been detected in early RM campaigns (e.g., Collier et al. \citeyear{collier_steps_1998}; Sergeev et al. \citeyear{sergeev_lagluminosity_2005}), it was not until intensive broadband reverberation mapping (IBRM) programs emerged that the time delays were rendered statistically significant (e.g., Edelson et al. \citeyear{edelson_space_2015}; Cackett et al. \citeyear{cackett_accretion_2018};  Hernández Santisteban et al. \citeyear{hernandezsantisteban_intensive_2020}; Kara et al. \citeyear{kara_agn_2021}). The improved quality (SNR\,$\sim$\,100) and cadence (<\,1\,day) of lightcurves allows for more robust testing of the standard geometrically thin accretion disc (Shakura \& Sunyaev \citeyear{shakura_black_1973}) and light reprocessing models. Early results from the IBRM campaigns revealed discrepancies between measured time delays and the expectation from our fundamental assumptions (e.g., Fausnaugh et al. \citeyear{fausnaugh_space_2016}), which prompted a myriad of alternative pictures for the disc structure and light interplay mechanisms (e.g., Gardner \& Done \citeyear{gardner_origin_2017}; Korista \& Goad \citeyear{korista_quantifying_2019}; Netzer \citeyear{netzer_testing_2020}; Sun et al. \citeyear{sun_corona-heated_2020}; Kammoun et al. \citeyear{kammoun_modelling_2021}; Starkey et al. \citeyear{starkey_rimmed_2022}; Hagen et al. \citeyear{hagen_what_2024}). IBRM campaigns are evidently crucial for deciphering AGN anatomy and revising the unified model (e.g., Urry \& Padovani \citeyear{urry_unified_1995}).

To achieve the necessary cadence during these IBRM campaigns, ground-based observatories combine data from multiple telescopes around the world, such as the 1-m robotic telescope network of the Las Cumbres Observatory (LCO; Brown et al. \citeyear{brown_cumbres_2013}), to give sub-daily observations in the optical. Combining observations from different telescopes requires careful intercalibration of the time-series data because of systematic differences in the instrumentation. Poor calibration can lead to telescope-specific offsets, seen as a `splitting' effect in the lightcurves (examples of such datasets can be seen in Donnan et al. \citeyear{donnan_testing_2023} and Hernández Santisteban et al. \citeyear{hernandezsantisteban_intensive_2020}), as well as underestimated uncertainties. This consequently induces artificial short timescale variability patterns not intrinsic to the AGN, which can lead to inaccurate time delay measurements when using lightcurve modelling tools or even simple cross-correlation analysis.

Several techniques are used to intercalibrate AGN lightcurves, by using iterative scaling (e.g., Hernández Santisteban et al. \citeyear{hernandezsantisteban_intensive_2020}) or modelling the AGN variability (e.g., {\tt PyCALI}, Li et al. \citeyear{li_bayesian_2014}; {\tt PyROA}, Donnan et al. \citeyear{donnan_bayesian_2021}). These methods can also include an additional variance parameter to account for underestimated uncertainties. Even though these tools are effective in reducing the telescope-specific offsets, the uncertainty estimation and outlier detection can be sub-optimal.

We present a new algorithm to intercalibrate AGN lightcurves across different telescopes which does not use the AGN itself but rather 10s to 100s of comparison stars on the same images, building upon the established technique of ensemble photometry (e.g., Honeycutt \citeyear{honeycutt_ccd_1992}; Fernández et al. \citeyear{fernandez_fernandez_improved_2012}). The stars suffer from the same epoch-specific and telescope-specific offsets, and the correction parameters derived from the algorithm are applied to the AGN data at the end. The algorithm simultaneously computes a noise model based on scatter in the star data, effectively identifying problematic epochs, telescopes, and stars. In Section\,\hyperlink{Section2}{2} we describe the mathematical background and the iterative algorithm. In Section\,\hyperlink{Section3}{3} we demonstrate the algorithm with LCO observations of NGC\,3783 and compare the results with the {\tt PyROA} intercalibration method, which uses the AGN lightcurve without reference to the comparison stars. 
In Section\,\hyperlink{Section4}{4} we discuss residual colour-dependent trends and
describe additional corrections that can be made for significantly bluer AGNs like Fairall\,9. 
We conclude with a brief summary in 
Section\,\hyperlink{Section5}{5}.

\hypertarget{Section2}{}

\section{Intercalibration Method}

\subsection{Mathematical Background}
Differential photometry measures the brightness of the variable AGN relative to non-variable comparison stars in an attempt to compensate for changes in sky transparency, airmass, sky background, and seeing. Comparing images taken with different telescopes, systematic offsets also arise due to differences in the camera sensitivities, filter transmission, and general calibration of each instrument. 
Instrumental star lightcurves therefore exhibit the same epoch-specific and telescope-specific systematic offsets as the AGN data. In this paper, the instrumental magnitude is defined as:
\begin{equation}
    m = -2.5\log(F_*),
\end{equation}
where $F_*$ is the background-subtracted count rate within the aperture in ADU/s.

The calibrated comparison star lightcurves should be constant, within some intrinsic scatter. 
Accordingly, the uncalibrated data are modelled as star-specific mean magnitudes ($\Bar{m}_\star$) offset by telescope-specific ($\Delta m_{\rm Tel}$) and epoch-specific ($\Delta m_{\rm Ep}$) magnitude corrections. 
\hypertarget{Eqmagcor}{}
\begin{equation} \label{eq:magcor}
    m(\star, \text{Tel}, \text{Ep}) = \bar{m}_\star + \Delta m_{\text{Tel}} + \Delta m_{\text{Ep}}\ .
\end{equation}
The corrections are multiplicative in fluxes but additive in magnitudes\footnote{Any epoch-dependent additive offsets in flux units are eliminated by adequate background subtraction and the use of large enough apertures in the photometry (see Hern\'andez Santisteban et al. \citeyear{hernandezsantisteban_intensive_2020} for a detailed description of how this is achieved for our data).}.
The degeneracy in the above model is removed by imposing a zero mean (or median) on the correction terms,
$
\left< \Delta m_{\text{Tel}} \right> =
\left< \Delta m_{\text{Ep}} \right>
= 0$.

By using 10s to 100s of comparison stars in each field, we can place tighter constraints on the
calibration parameters than with standard differential photometry using typically just a few comparison stars to determine such offsets. 
Additionally, our algorithm allows for different numbers of comparison stars at different epochs, 
which can occur due to loss of transparency or to pointing changes that move comparison stars outside the detector field of view. 
Honeycutt (\citeyear{honeycutt_ccd_1992}) describes in detail the use of such ensemble photometry to derive epoch-specific corrections for inhomogeneous datasets\footnote{A {\tt Python} implementation of this method is available at \url{https://github.com/connorrobinson/ensemble}.}, but the additional telescope-specific corrections are not considered.

\begin{figure}
    \centering
    \includegraphics[width=\columnwidth]{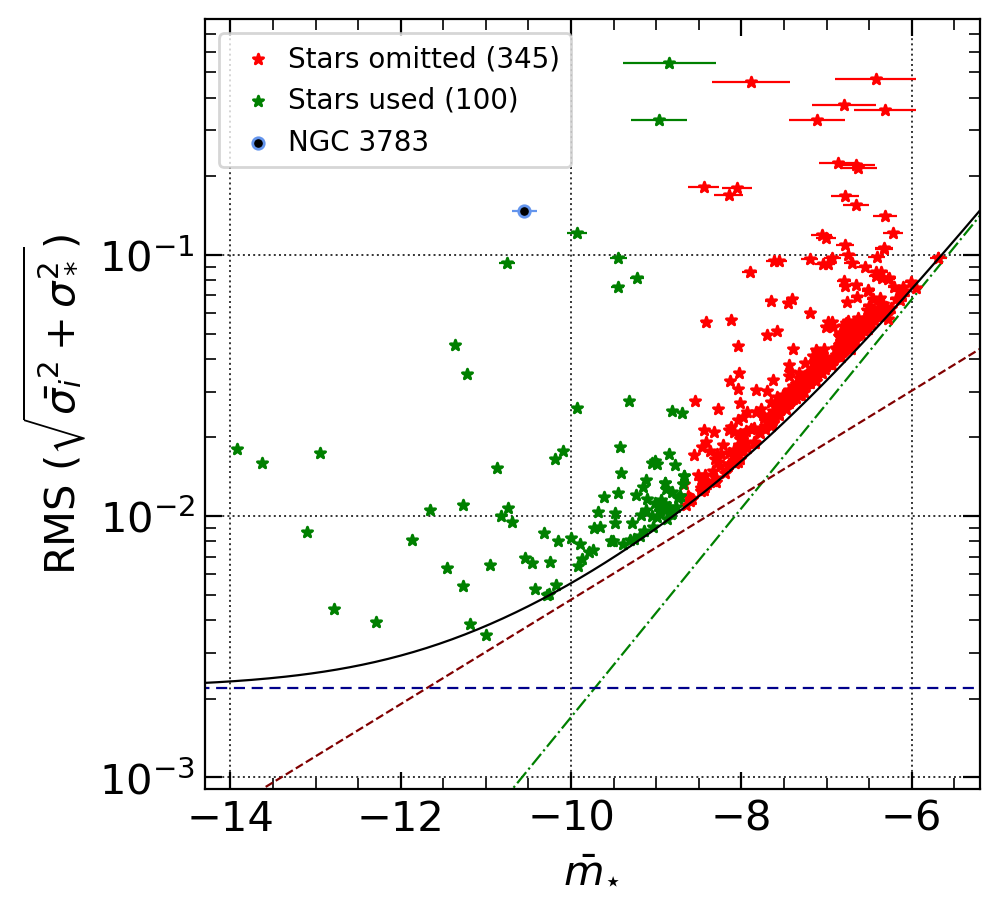} 
    \caption{  \label{fig:rmsmag}
    RMS as a function of $\bar{m}_{\star}$ for 445 field stars around NGC\,3783 in the LCO \textit{gp} band, where $\bar{m}_{\star}$ is the mean instrumental star magnitude derived from our algorithm. 
    The hundred brightest stars (green) are selected for the intercalibration. 
    The RMS of each star shown here is the mean of the nominal uncertainties ($\bar{\sigma}_i^2 = \frac{1}{N_{\text{Ep}}} \sum_{i} \sigma_i^{2}$) added in quadrature with the star-specific extra variance parameter derived from our algorithm. 
    The black curve is the square sum of the star Poisson noise (red short-dashed), background noise (green dot-dashed), and the constant scale noise (blue long-dashed),  approximating the lower limit for the expected RMS as a function of star magnitude. The higher RMS at $\bar{m}_{\star} \lesssim-12$ is due to CCD saturation. The RMS of the AGN is also shown (black point), demonstrating its highly variable nature. The AGN has $N_{\text{Ep}}$ = 1119 in the LCO \textit{gp} band.
    }
\end{figure}

We introduce a multi-component noise model that accounts for stellar variability and underestimated uncertainties:
\hypertarget{Eqvar}{}
\begin{equation} \label{eq:var}
    \sigma^{2}(\star\text{, Tel, Ep}) = \sigma^{2}_{i} + \sigma^{2}_\star + \sigma^{2}_{\rm Tel} + \sigma^{2}_{\rm Ep} \ ,
\end{equation}
where $\sigma_{i}^2$ is the nominal variance,
$\sigma_{\rm \star}^2$ is a star-specific extra variance parameter,
$\sigma_{\rm Tel}^2$ is a telescope-specific extra variance parameter, 
and $\sigma_{\rm Ep}^2$ is an epoch-specific extra variance parameter.
This model assumes no covariance among the noise model parameters.
Stars that are intrinsically more variable than others 
introduce a larger scatter in the correction parameters since our model in Eq.\,(\hyperlink{Eqmagcor}{\ref{eq:magcor}}) assumes all stars are of constant magnitude, and thus by including $\sigma_\star^2$ the contribution from variable stars is down-weighted. 
An RMS versus mean magnitude plot for all the comparison stars reveals intrinsically variable stars as those that sit significantly above the expected variance for a given magnitude in different noise regimes (Fig\,\hyperlink{Figrmsmag}{\ref{fig:rmsmag}}). 
These variable stars can be omitted in the calibration if the number of comparison stars is limited.

With lightcurve data on $N_\star$ stars at $N_{\rm Ep}$ epochs by $N_{\rm Tel}$ telescopes,
the calibration model has $N_{\rm P} = 2\,\left(N_\star+ \left(N_{\rm Ep}-1\right)+\left(N_{\rm Tel}-1\right)\right)$ parameters constrained by $N_{\rm D}=N_\star\,N_{\rm Ep}$ data.
The calibration is thus well constrained when $N_{\rm D} > N_{\rm P}$.

We estimate the $N_{\rm P}$ calibration
parameters by fitting all $N_{\rm D}$ data from the uncalibrated comparison star lightcurves. 
To optimise each set of calibration parameters\footnote{`Each set' ($\mu$, $\sigma_{\text{x}}^2$) refers to either the epoch-specific ($\Delta m_{\rm Ep}$, $\sigma_{\rm Ep}^2$), telescope-specific ($\Delta m_{\rm Tel}$, $\sigma_{\rm Tel}^2$), or star-specific ($\Bar{m}_\star$, $\sigma_{\rm *}^2$) model parameters.} ($\mu$, $\sigma_{\text{x}}^{2}$), when presented with a scatter of $N$ magnitude residuals
($x_{i}\pm\sigma'_i$) from the relevant epochs, telescopes and comparison stars, we 
 maximise the likelihood function $\mathcal{L}$, 
 for Gaussian errors, equivalent to minimising the Badness of Fit (BoF) statistic:
\hypertarget{Eqbof}{}
\begin{equation} \label{eq:bof}
    -2\ln \mathcal{L} = \sum\limits_{i=1}^N \frac{(x_{i} - \mu)^{2}}{\sigma^2_{\rm x} + (\sigma'_i)^2} +\sum\limits_{i}^N \ln\left[\sigma_{\rm x}^2 + (\sigma'_i)^2\right] + N\,\ln{2\pi}\ .
\end{equation}
Here the $x_{i}$ are magnitude residuals, $\mu$ is the desired magnitude correction parameter or mean star magnitude,  $\sigma_{\rm x}^{2}$ is the corresponding extra variance parameter, and $(\sigma'_i)^2$ is the noise model (Eq.\,\hyperlink{Eqvar}{\ref{eq:var}}) omitting the desired extra variance parameter.
The first term in Eq.\,(\hyperlink{Eqbof}{\ref{eq:bof}}) is the chi-squared statistic, with the extra variance parameter in the denominator added in quadrature to the noise model. 
The second term is the penalty that prevents the extra variance parameter from minimising $\chi^2$ by increasing it to arbitrarily large values ($\sigma_{\rm x}\rightarrow \infty$).

To minimise the BoF, partial derivatives of Eq.\,(\hyperlink{Eqbof}{\ref{eq:bof}}) with respect to 
$\mu$ and $\sigma_{\rm x}^2$ must vanish:
\hypertarget{Eqbofmu}{}
\begin{equation} \label{eq:bofmu}
    0 = \frac{\partial (-2\ln \mathcal{L})}{\partial \mu} = -2\sum_{i = 1}^{N}\frac{x_{i} - \mu}{\sigma_{\rm x}^2 + (\sigma'_i)^2} \ ,
\end{equation}
\hypertarget{Eqbofvar}{}
\begin{equation} \label{eq:bofvar}
 0 = \frac{\partial (-2\ln\mathcal{L})}{\partial \sigma_{\rm x}} = -\sum_{i = 1}^{N} 
 \left( \frac{(x_{i} - \mu)^{2}}{[\sigma_{\rm x}^2 + (\sigma'_i)^2]^{2}} + 
 \frac{1}{\sigma_{\rm x}^2 + (\sigma'_i)^2} \right) \ .
\end{equation}
Introducing $g_{i} \equiv \sigma_{\text{x}}^2/\left(\sigma_{\rm x}^2 + (\sigma'_i)^2 \right)$, and then rearranging to solve for $\hat{\mu}$ and $\hat{\sigma}_{\rm x}^2$,
the optimal parameter estimates are:
\hypertarget{Eqmu}{}
\begin{equation} \label{eq:mu}
    \hat{\mu}(\sigma_{\rm x}) = \frac{\sum_i x_{i}\,g_{i}}{\sum_i g_{i}}
    \ ,
\end{equation}

\hypertarget{Eqvarx}{}
\begin{equation} \label{eq:varx}
    \hat{\sigma}_{\text{x}}^{2}(\mu) = \frac{\sum_i (x_{i} - \mu)^{2}\,g_i^2}{\sum_i g_i}
    \ .
\end{equation}
Note that $0<g_i<1$. We refer to $g_i$ as the `goodness' of datum $i$ for provision of information on $\sigma_{\rm x}$.
Because $\hat{\mu}$ depends on $\hat{\sigma}_{\rm x}$, and vice versa,
Eq.\,(\hyperlink{Eqmu}{\ref{eq:mu}}) and (\hyperlink{Eqvarx}{\ref{eq:varx}}) are solved iteratively until a convergence is reached. 
To do this we refer to the code {\tt arx}\footnote{\url{https://github.com/Alymantara/arx}}, which takes in $N$ data values $x_{i}$ with uncertainties $\sigma'_i$, and returns maximum-likelihood estimates for their mean (Eq.\,\hyperlink{Eqmu}{\ref{eq:mu}}) and extra variance (Eq.\,\hyperlink{Eqvarx}{\ref{eq:varx}}).

\begin{figure*}
    \centering
    \includegraphics[width=\textwidth]{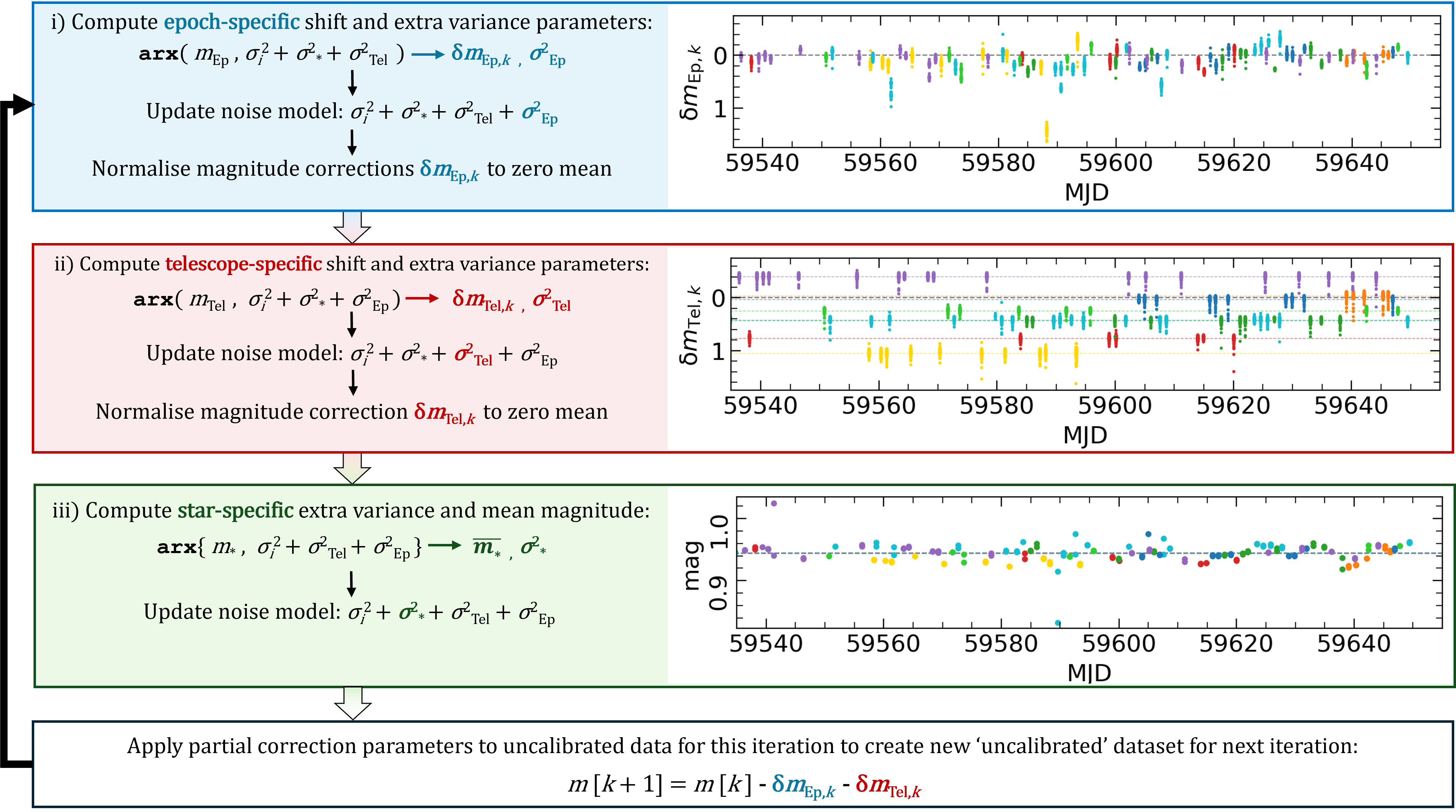} 
    \caption{ 
    \label{fig:pytics}
    Schematic of the iterative intercalibration algorithm {\tt PyTICS}, demonstrated on a section of NGC\,3783 data for the LCO \textit{up} band. Datapoints are coloured by telescope. 
    \textit{Top}: Datapoints are grouped by epoch (coloured group at each MJD) after subtracting the mean telescope magnitude for each star. 
    The optimal values of $\delta m_{\rm Ep}$ and $\sigma_{\rm Ep}^2$ are computed given each epoch residual cluster and the outlined noise model.
    \textit{Middle}: Datapoints are grouped by telescope (coloured rows and dashed lines showing the mean for each telescope) after correcting the uncalibrated data with $\delta m_{\rm Ep}$ and subtracting the mean star magnitude for each star.
    The optimal values of $\delta m_{\rm Tel}$ and $\sigma_{\rm Tel}^2$ are computed given each telescope residual cluster and the outlined noise model.
    \textit{Bottom}: The uncalibrated data of each star is corrected with $\delta m_{\rm Ep}$ and $\delta m_{\rm Tel}$. The optimal values of $\Bar{m}_\star$ and $\sigma_\star^2$ are computed given all datapoints for each star and the outlined noise model. At the end of each iteration, the next `uncalibrated' dataset is constructed by applying the partial correction parameters and updating the total noise model. Note that the datapoints here are shown without errorbars.}
\end{figure*}

The uncertainties for $\hat{\mu}$ and $\hat{\sigma}_{\rm x}^2$ are estimated by considering the second derivative of the BoF with respect to the desired parameters\footnote{This approximation for the variance of the model parameters originates from the Taylor expansion of the BoF statistic around the minimum at $\hat{\mu}$ or $\hat{\sigma}_{\rm x}^2$. This method removes the covariance between parameters.}:
\hypertarget{Eqvarmu}{}
\begin{equation} \label{eq:varmu}
    \text{Var}(\hat{\mu}) \approx \frac{2}{\frac{\partial^{2} (-2\ln\mathcal{L})}{\partial \mu ^{2}}} 
    = \frac{{\sigma}_{\text{x}}^{2}}{\sum_i g_i} \ ,
\end{equation}
\hypertarget{Eqvarvarx}{}
\begin{equation} \label{eq:varvarx}
    \text{Var}(\hat{\sigma}_{\text{x}}^{2}) \approx \frac{2}{\frac{\partial^{2} (-2\ln{\mathcal{L}})}{\partial (\sigma_{\rm x}^2)^{2}}} 
    = \frac{2\,\sigma_{\rm x}^{6}}
    {2\sum_i(x_{i}-\mu)^{2}g_{i}^3 
    - \sigma_{\rm x}^{2}\sum_i{g_{i}^{2}}} \ . 
\end{equation}
The parameter uncertainties in Eq.\,(\hyperlink{Eqvarmu}{\ref{eq:varmu}}) and Eq.\,(\hyperlink{Eqvarvarx}{\ref{eq:varvarx}}) are also computed in {\tt arx}, where they are used to track the convergence.

\subsection{Iterative Optimal Scaling}

Fig.\,\hyperlink{Figpytics}{\ref{fig:pytics}} illustrates the iterative optimal scaling algorithm, highlighting the arguments given to {\tt arx} when computing each set of optimal parameters. We assume initially that the extra variance parameters ($\sigma_\star^2$, $\sigma_{\rm Tel}^2$, $\sigma_{\rm Ep}^2$) and magnitude shifts ($\Delta{m}_{\rm Tel}$, $\Delta{m}_{\rm Ep}$) all vanish, and update the values throughout each iteration. During each iteration $k$, we compute the \textit{partial} correction parameters $\delta m_{\text{Tel}, k}$ and $\delta m_{\text{Ep}, k}$, which are summed to produce the final correction parameters $\Delta m_{\rm Tel}$ and $\Delta m_{\rm Ep}$ after $M$ iterations:
\hypertarget{Eqdmep}{}
\begin{equation} \label{eq:dmep}
    \Delta m_{\rm Ep} = \sum_{k = 1}^{M} \delta m_{\text{Ep}, k} \ ,
\end{equation}
\hypertarget{Eqdmtel}{}
\begin{equation} \label{eq:dmtel}
    \Delta m_{\rm Tel} = \sum_{k = 1}^{M} \delta m_{\text{Tel}, k} \ .
\end{equation}

The iterative intercalibration algorithm is as follows:

\begin{enumerate}
    \item Compute epoch-specific corrections, $\delta m_{\text{Ep}, k}$. For each star, we separate the data by telescope and compute the optimal average\footnote{The optimal average is the inverse-variance weighted average $\hat{X} = \frac{\sum{w_{j}x_{j}}}{\sum w_{j}}$, where $w_{j} = \frac{1}{\sigma_{j}^2}$.}.The residuals between this telescope mean and the corresponding magnitude data are the epoch-specific offset parameters $\delta m_{\text{Ep}, k}$ for this iteration (Fig.\,\hyperlink{Figpytics}{\ref{fig:pytics}} top). The optimal value and extra variance for each epoch-specific correction is obtained with {\tt arx} given the residuals from all comparison stars as previously described. The total noise model is updated with the new $\sigma_{\rm Ep}^2$, and the magnitude corrections $\delta m_{\text{Ep}, k}$ are normalised to zero mean to mitigate the degeneracy in the uncalibrated data model.
    \item Compute telescope-dependent corrections, $\delta m_{\text{Tel}, k}$. For each star, the newly computed correction parameters $\delta m_{\text{Ep}, k}$ are subtracted from the uncalibrated lightcurves, reducing the epoch-specific shifts in the data. The optimal average is then computed using all datapoints to obtain a mean star magnitude. The residuals between this star mean and the corresponding telescope magnitude data are the telescope-dependent offset parameters $\delta m_{\text{Tel}, k}$ for this iteration (Fig.\,\hyperlink{Figpytics}{\ref{fig:pytics}} middle). The optimal value and extra variance for each telescope-dependent correction is obtained with {\tt arx} given the residuals from all comparison stars as previously described. The total noise model is updated with the new $\sigma_{\rm Tel}^2$, and the magnitude corrections $\delta m_{\text{Tel}, k}$ are normalised to zero mean to mitigate the degeneracy in the uncalibrated data model. 
    \item Compute $\Bar{m}_\star$. For each star, having subtracted the newly computed correction parameters $\delta m_{\text{Tel}, k}$ and $\delta m_{\text{Ep}, k}$, the datapoints are brought closer to the mean magnitude of the star (Fig.\,\hyperlink{Figpytics}{\ref{fig:pytics}} bottom). The optimal value and extra variance is obtained with {\tt arx} given the corrected magnitude data for each star as previously described. The noise model is updated with the new $\sigma_\star^{2}$. The newly corrected star lightcurves during this step are used as the new `uncalibrated' dataset for the next iteration. Convergence is reached when 
    all parameters change between iterations by less than a small fraction (e.g. $10^{-5}$) of the corresponding parameter uncertainties.
\end{enumerate}

The total epoch-specific and telescope-specific correction parameters (Eq.\,\hyperlink{Eqdmep}{\ref{eq:dmep}} and Eq.\,\hyperlink{Eqdmtel}{\ref{eq:dmtel}}, respectively) are applied to the uncalibrated AGN lightcurve at the end\footnote{Here we are excluding the AGN itself from the ensemble of stars used in the iterative algorithm. In principle the AGN can be included in the ensemble and will be registered as a `variable star' that is down-weighted with a relatively large $\sigma_\star^{2}$ parameter. The extra variance parameter found for the AGN lightcurve data measures the RMS amplitude of its intrinsic variations (see Fig\,\hyperlink{Figrmsmag}{\ref{fig:rmsmag}}).}. The nominal uncertainties are added in quadrature with the telescope-specific and epoch-specific extra variance parameters. The star-specific extra variance parameters quantify the variability amplitude of each star, and thus serve to down-weight the contribution of intrinsically more variable stars during the iterative process. Large gaps in the data, such as seasonal gaps, are no issues for the algorithm as the AGN lightcurve does not need to be interpolated, and undersampled parts of the lightcurve have no effect on the errorbar estimation.

We refer to our algorithm as {\tt PyTICS} (\textbf{T}elescope \textbf{I}ntercalibration using \textbf{C}omparison \textbf{S}tars) for the remainder of this paper.

\begin{figure*}
    \centering
    \includegraphics[width=\textwidth]{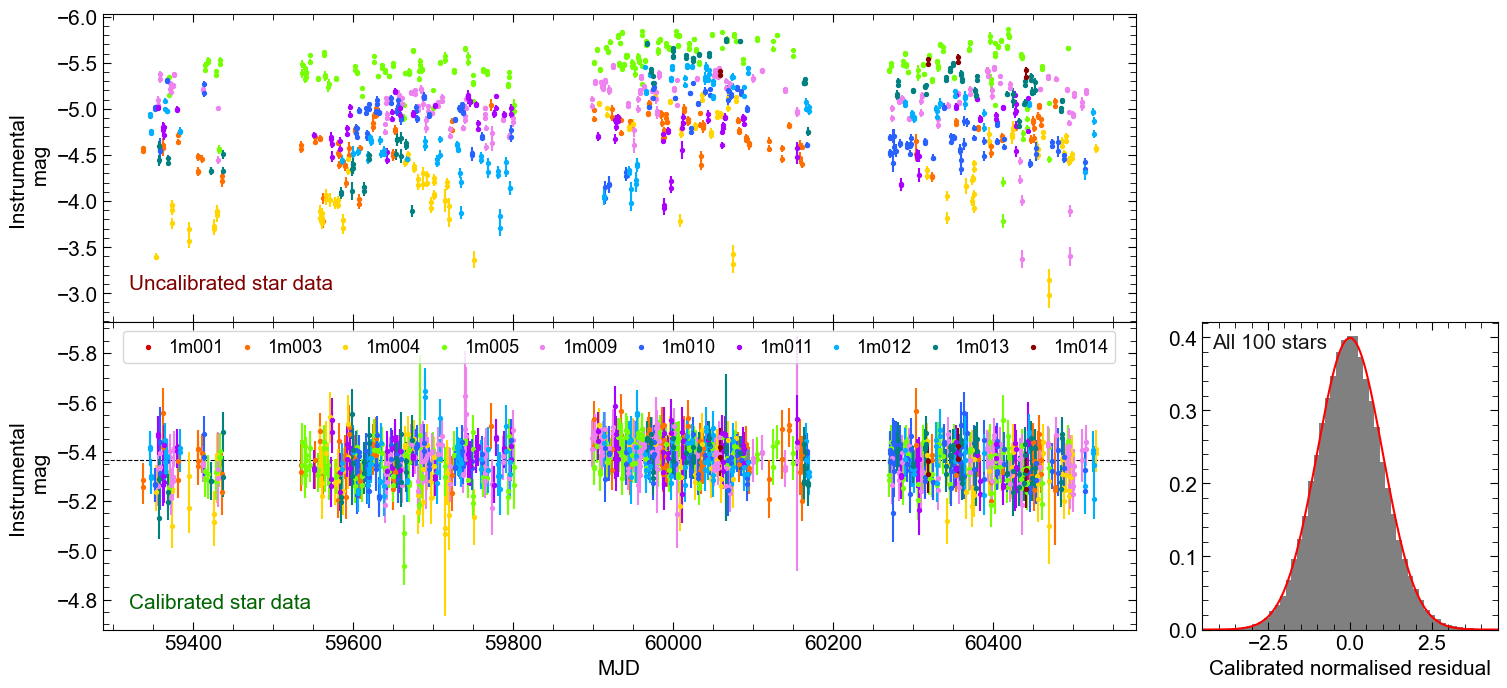} 
    \caption{ \label{fig:starcal}
    \textit{Top}: Uncalibrated instrumental magnitude lightcurve for a non-variable star in the field of NGC 3783, for the LCO \textit{ip} band. 
    The datapoints are coloured according to the telescopes in the LCO network, and show significant offsets as well as errorbars not representative of the scatter. 
    \textit{Bottom}: Intercalibrated lightcurve using {\tt PyTICS}. 
    The correction parameters derived from the algorithm bring the data to a mean magnitude, with scatter appropriately described by the noise model. \textit{Right}: The normalised residuals for all 100 stars used in the calibration (gray), showing a standard Gaussian distribution ($\mu$ = 0, $\sigma = 1$) in red.
    }
\end{figure*}

\begin{figure*}
    \centering
    \includegraphics[width=\textwidth]{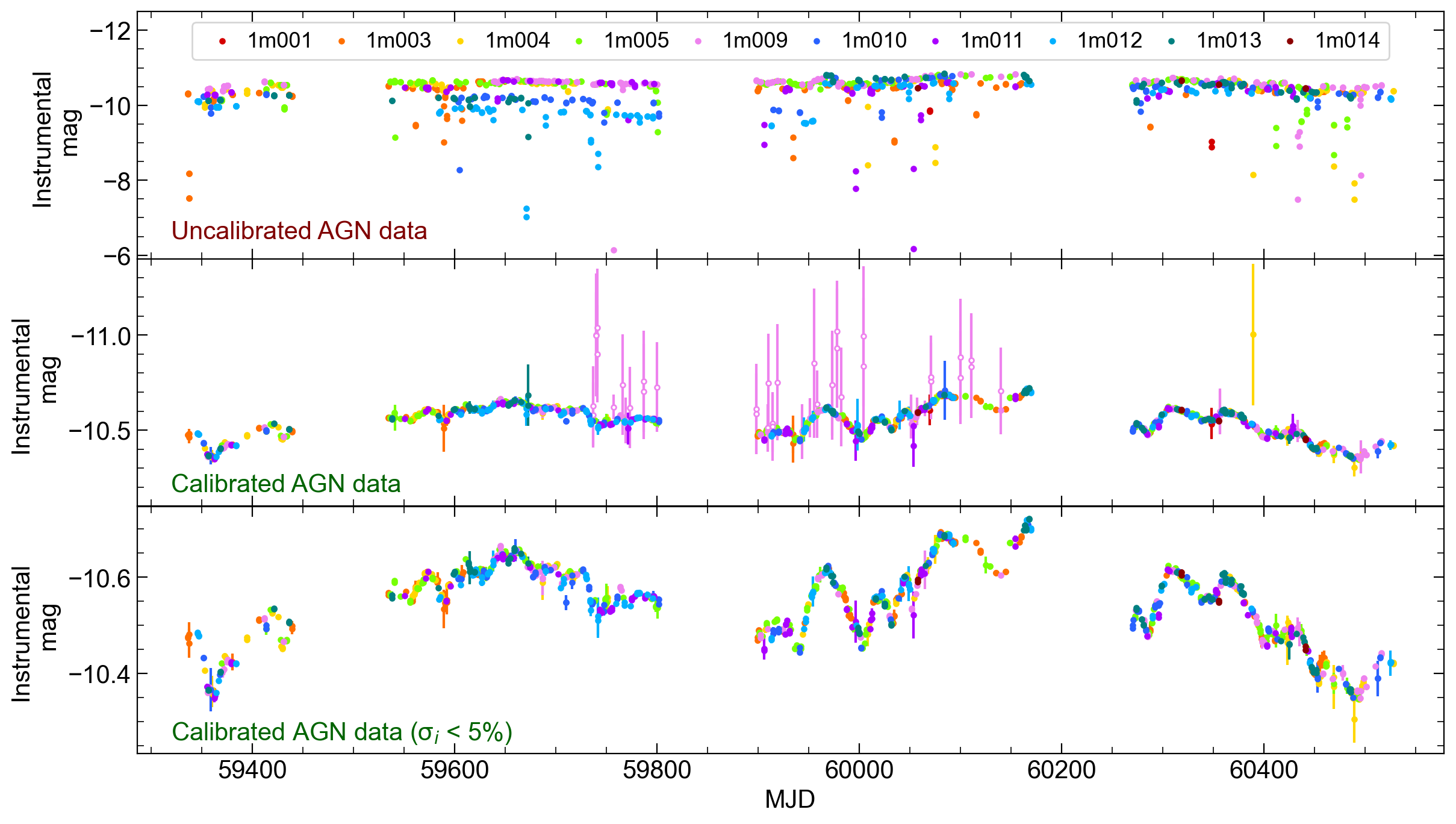} 
    \caption{  \label{fig:agncal}
    \textit{Top}: Uncalibrated instrumental magnitude lightcurve for NGC\,3783 in the LCO \textit{ip} band, with datapoints coloured according to the telescope. 
    \textit{Middle}: Intercalibrated lightcurve using {\tt PyTICS}. 
    The variability pattern is more representative of AGN variations. 
    The extreme outliers (hollow pink markers) are from a period of time when telescope 1m009 was experiencing focus issues. 
    \textit{Bottom}: Same as above, but with data clipped below 5 percent uncertainty.
    }
\end{figure*}

\section{Intercalibrating observations}

\hypertarget{Section3}{}

\subsection{NGC 3783}

We demonstrate the intercalibration algorithm on photometric lightcurves of the nearby ($z$ = 0.00973, Theureau et al. {\citeyear{Theureau1998}}) Seyfert I galaxy NGC\,3783, using data from ten LCO 1-m telescopes across four sites (see Table~\ref{tab:my_label} for details). The {\tt PyTICS} algorithm was tested on all 7 of the LCO bands in our database. The observations were taken as part of the LCO Key Projects  KEY2020B-006 and KEY2023B-001 (P.I. Hernández Santisteban) from MJD 59336 to 60527 with approximately daily cadence. A full description of the LCO data reduction can be found in Hern\'andez Santisteban (in prep).
\begin{table*}
    \caption{LCO telescope network information and NGC\,3783 \textit{B} band data.}
    \centering
    \begin{tabular}
    {|c|c|c|c|c|c|}
        \multicolumn{2}{c|}{\textbf{LCO Telescope Information}} & \multicolumn{4}{c}{\textbf{NGC\,3783 \textit{B} Band Data}} \\
        \hline
        \textbf{ID} & \textbf{Site} & \textbf{Mean Epoch \#} & \textbf{$\Delta m_{\text{Tel}}$}& \textbf{$\sigma_{\text{Tel}}$} &
        \textbf{\textit{B-V} Colour-Residual Slope}\\
        \hline
        1m004 & Cerro Tololo Interamerican Observatory, Chile & 141.9 & 0.3485 $\pm$ 0.00013 & 0.0074 & -0.0454 $\pm$ 0.0042 \\
        1m005 & Cerro Tololo Interamerican Observatory, Chile & 261.7 & -0.2160 $\pm$ 0.00005 & 9.6$\times10^{-5}$ & 0.0441 $\pm$ 0.0021 \\
        1m009 & Cerro Tololo Interamerican Observatory, Chile & 192.5 & -0.0009 $\pm$ 0.00007 & 0.0026 & -0.0063 $\pm$ 0.0032 \\
        1m010 & South African Astronomical Observatory, South Africa & 154.4 & 0.4039 $\pm$ 0.00011 & 0.0057 & 0.0237 $\pm$ 0.0037 \\
        1m012 & South African Astronomical Observatory, South Africa & 141.6 & 0.3372 $\pm$ 0.00009 & 0.0004& -0.0028 $\pm$ 0.0025 \\
        1m013 & South African Astronomical Observatory, South Africa & 110.7 & 0.1505 $\pm$ 0.00010 & 0.0009 & 0.0418 $\pm$ 0.0018 \\
        1m003 & Siding Spring Observatory, Australia & 160.6 & 0.1838 $\pm$ 0.00010 & 0.0044 & -0.0519 $\pm$ 0.0023 \\
        1m011 & Siding Spring Observatory, Australia & 114.7 & 0.0762 $\pm$ 0.00011 & 0.0041 &-0.0511 $\pm$ 0.0022 \\
        1m001 & Teide Observatory, Spain & 1.7 & 1.3404 $\pm$ 0.01232 & 0.1546 & -0.0875 $\pm$ 0.0906 \\
        1m014 & Teide Observatory, Spain & 7.9 & -0.1535 $\pm$ 0.00050 & 0.0085 & -0.0089 $\pm$ 0.0057 \\
        1m006 & McDonald Observatory, Texas & 0 & & &  \\
        1m008 & McDonald Observatory, Texas & 0 & & &  \\
        \hline
        \multicolumn{6}{|p{\dimexpr\textwidth-2\tabcolsep}|}{\small \textit{Notes:} (1) LCO telescope ID; (2) Observatory location; (3) Mean number of epochs for each telescope across the 100 stars used in the calibration of NGC\,3783 for the LCO $B$ band; (4) The telescope-specific correction parameter, and its uncertainty, derived from the {\tt PyTICS} algorithm; (5) The telescope-specific extra variance parameter derived from the {\tt PyTICS} algorithm (6) The colour-residual slope, and its uncertainty, for the \textit{B - V} colour index, derived from NGC\,3783 data only.} \\
    \end{tabular}
    \label{tab:my_label}
\end{table*}

For the LCO \textit{ip} band, 430 comparison stars are available in the field. 
We omit stars with less than 50$\%$ of the maximum number of available datapoints from the ensemble. This leaves an ensemble of 303 comparison stars. 
Next, we arbitrarily choose 100 of the brightest remaining stars; using more than this does not significantly affect the final parameter estimates we derive from the algorithm. 
Choosing to omit fainter stars in the ensemble saves on compute time and avoids the regime where background noise begins to dominate over the photon counting noise.
Note that we include a handful of the brightest stars that suffer from saturation at some epochs (see Fig\,\hyperlink{Figrmsmag}{\ref{fig:rmsmag}}). 
As previously emphasised, inadvertently including variable (or saturated) stars in the ensemble is largely harmless because they are down-weighted by their star-specific extra variance parameters. 
This feature of the algorithm can be advantageous by avoiding the need to arbitrarily fine-pick a select subset of comparison stars.

Fig.\,\hyperlink{Figstarcal}{\ref{fig:starcal}} (top) shows the uncalibrated lightcurve for one comparison star, where the telescope-specific magnitude shifts are apparent, and the scatter is poorly represented by the nominal uncertainties. The second panel of Fig.\,\hyperlink{Figstarcal}{\ref{fig:starcal}} shows the calibrated lightcurve after 48 iterations, when all parameters changed by less than a factor of $10^{-5}$ of their estimated uncertainties. 
Qualitatively, the scatter in the data is well represented by the noise model, but to further quantify the quality of the calibration, we examine the normalized residuals.
Uncertainties that appropriately describe the scatter are expected to show a Gaussian distribution, which can be seen clearly from the histogram in Fig.\,\hyperlink{Figstarcal}{\ref{fig:starcal}} when examining all 100 stars.

In Fig.\,\hyperlink{Figagncal}{\ref{fig:agncal}} we show the result of {\tt PyTICS} for the AGN data, before (top) and after (middle) applying the correction parameters computed from the comparison stars. 
The typical AGN variability pattern is a lot clearer compared to the uncalibrated data, with no apparent telescope-specific shifts in the lightcurves. 
A notable exception is the 1m009 telescope, where some data are shifted by $\sim$0.3 magnitudes with larger scatter, and other data are close to the lightcurve defined by the other telescopes. 
After inspecting the images, we find that this telescope experienced focus issues during this period, causing blurred, doughnut-shaped stars. This focus issue was seen in images for other AGNs as well. 
The {\tt PyTICS} algorithm effectively uses the large scatter in the comparison star fluxes to quantify the problem, and assigns appropriately increased uncertainties to fluxes at the problematic epochs.

\subsection{Comparison of {\tt PyTICS} with {\tt PyROA}.}

Intercalibration of the AGN lightcurve is performed at the end of the iterative process, requiring no assumptions to be made about the variability shape, unlike in the RM data calibration software {\tt PyCALI} (Li et al. \citeyear{li_bayesian_2014}) for example. As has been widely the case in literature, the underlying variability is assumed to be a damped random walk (DRW; e.g., Kelly et al. \citeyear{kelly_are_2009}), however there is still ongoing discussion about the validity of this model 
(e.g., Kasliwal et al. \citeyear{kasliwal_are_2015}; Simm et al. \citeyear{simm_pan-starrs1_2016}). 

\begin{figure}
    \centering
    \includegraphics[width=\columnwidth]{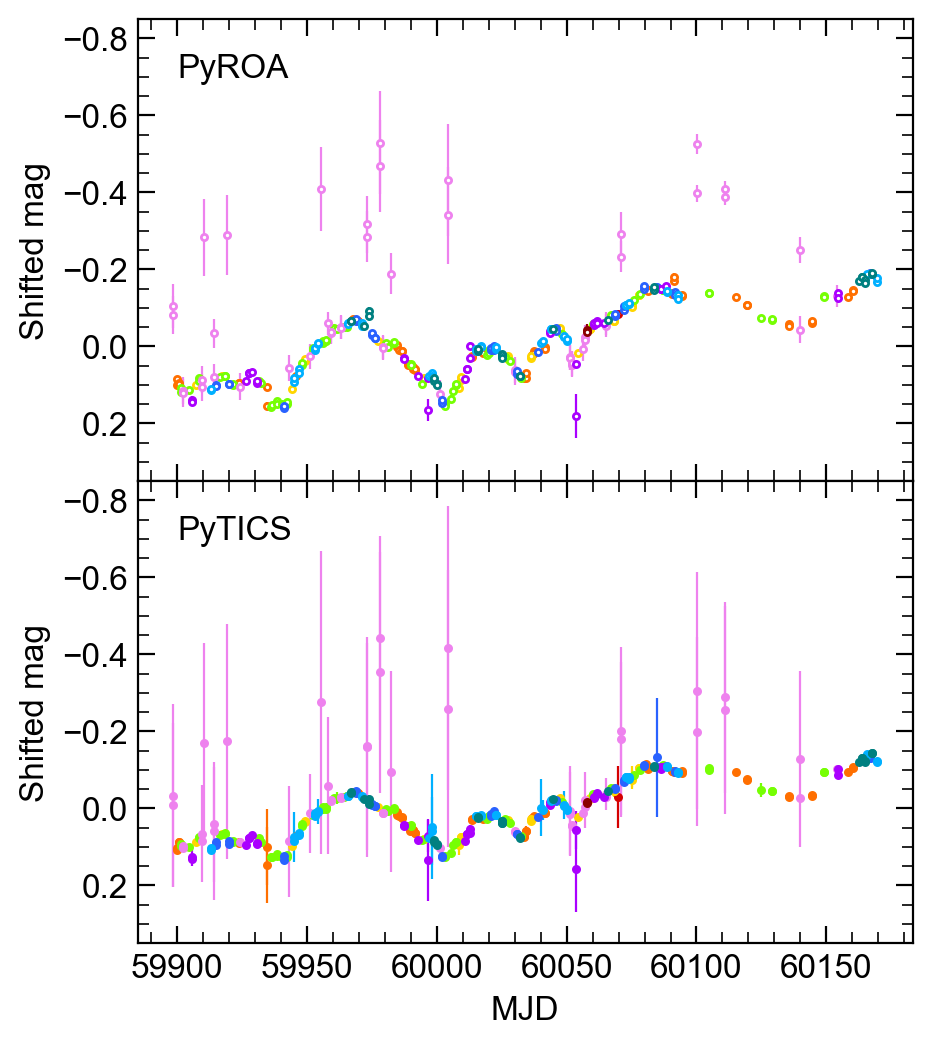} 
    \caption{
    \label{fig:pyroa}
    A section of the NGC\,3783 lightcurve for the LCO \textit{ip} band, intercalibrated using {\tt PyROA} (\textit{top}) and {\tt PyTICS} (\textit{bottom}). Datapoints are coloured according to the telescope. The large pink outliers are from telescope 1m009, which experienced focus issues during this period.
    }
\end{figure}

{\tt PyROA} (Donnan et al. \citeyear{donnan_bayesian_2021}) is another lightcurve modelling tool used to measure time delays between multi-wavelength observations, but also has the ability for intercalibration of photometric data. It models the lightcurves with a running optimal average (ROA) and uses MCMC sampling to determine the best fit telescope-specific scale factors and additive background flux offsets for intercalibrating the data. 
 The software also computes telescope-specific extra variance parameters by including an additional noise term in the log-likelihood function, similar to our approach in Eq.\,(\hyperlink{Eqbof}{\ref{eq:bof}}).
 Moreover, {\tt PyROA} performs a soft sigma clipping, where outlier uncertainties are expanded up to a 3$\sigma$ level. 
 The top panel in Fig.\,\hyperlink{Figpyroa}{\ref{fig:pyroa}} shows a section of the AGN lightcurve calibrated with {\tt PyROA}, demonstrating how datapoints further from the AGN lightcurve have increasingly larger uncertainties to remain capped at the chosen $\sigma$ threshold. The outlier detection is sub-optimal for datapoints in large gaps, however, as seen near MJD 60100, where the uncertainties are poorly describing the large deviation from the lightcurve.

\begin{figure}
\includegraphics[width=\columnwidth]{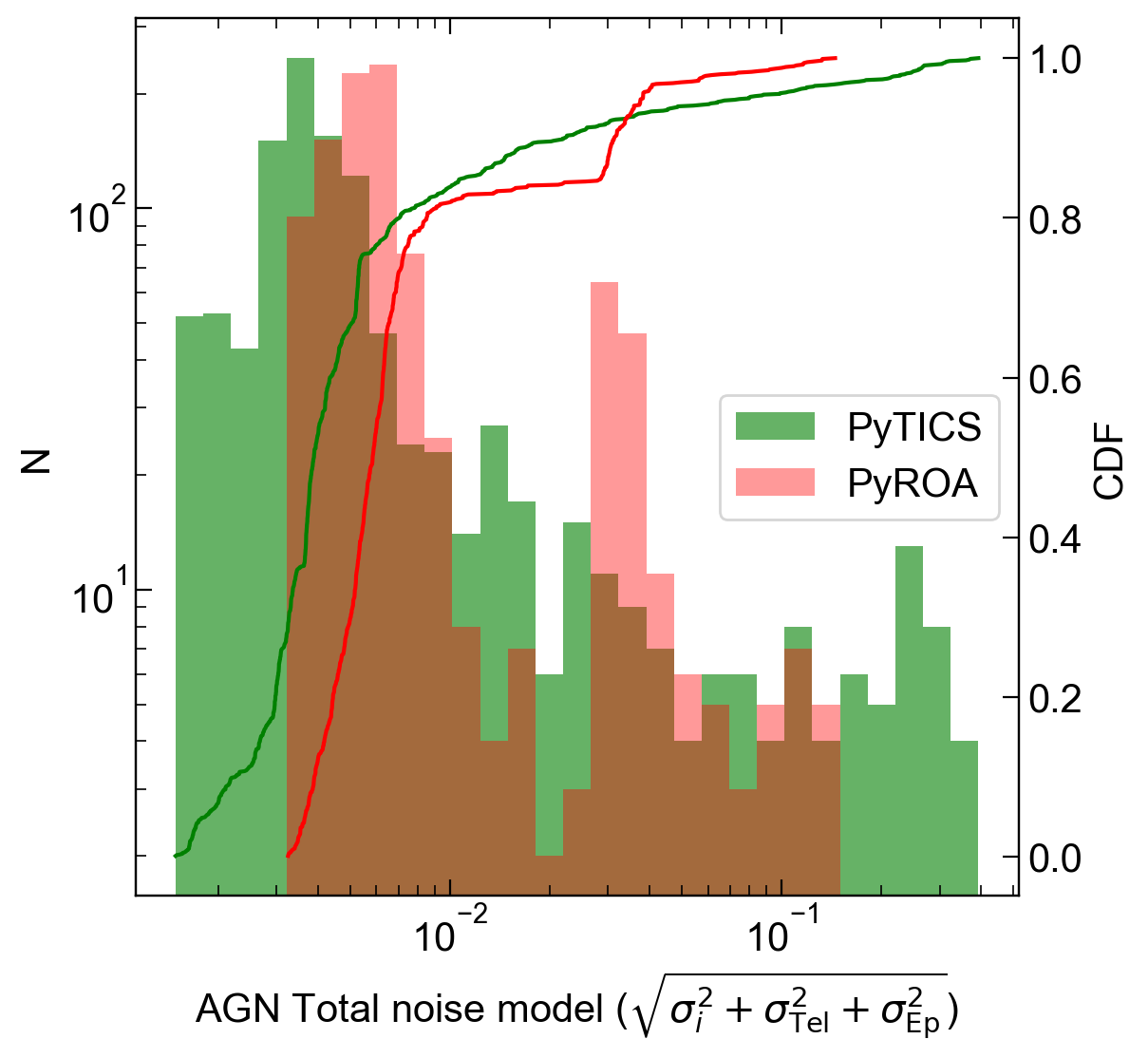} 
    \caption{
        \label{fig:hist}
        Distribution of the total noise model in magnitudes for NGC\,3783 in the LCO \textit{ip} band, derived from {\tt PyTICS} (green) and {\tt PyROA} (red), together with the cumulative distribution functions (CDF).
    }
\end{figure}


\begin{figure*}
    \centering
    \includegraphics[width=\textwidth]{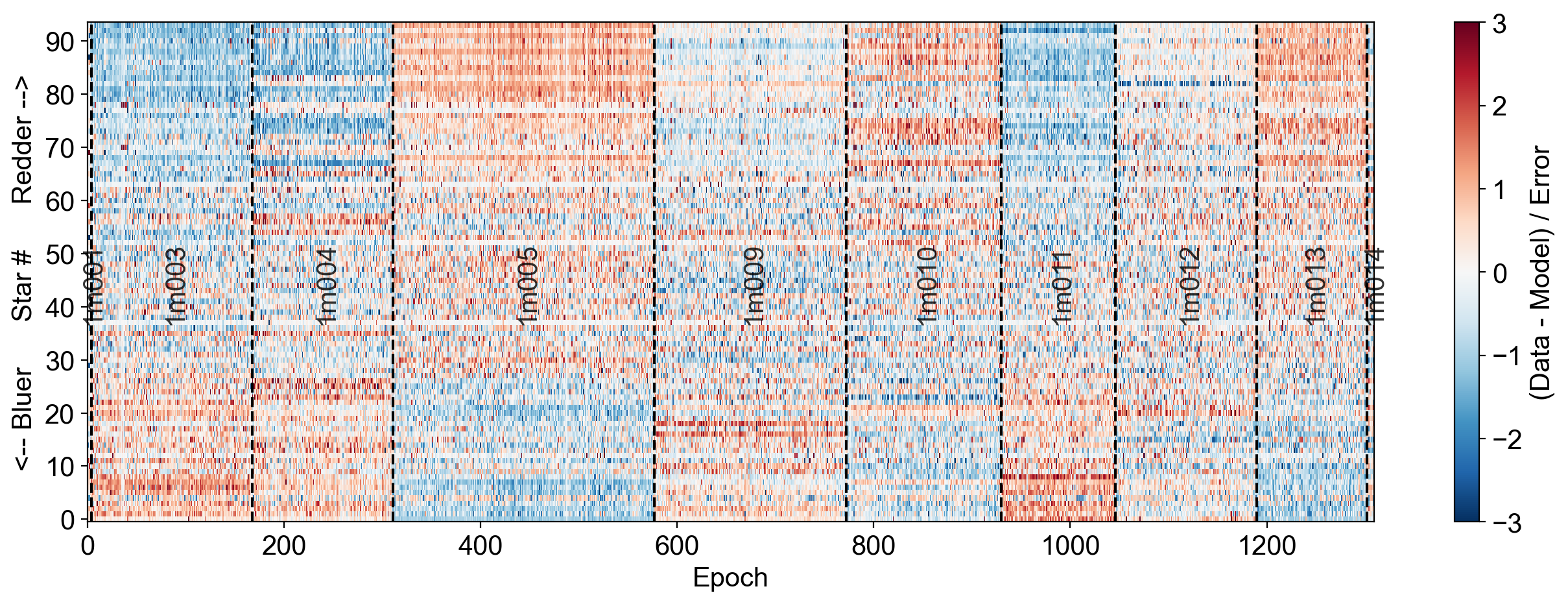} 
    \caption{ 
    \label{fig:heatmap}
    Normalised residual heatmap of the comparison stars used for the calibration of NGC\,3783 in the LCO \textit{B} band. 
    Epochs are sorted by telescope and stars are sorted by $B - V$ colour, showing clear residual trends with different slopes for each telescope.
    }
\end{figure*}

\begin{figure}
    \centering
    \includegraphics[width=\columnwidth]{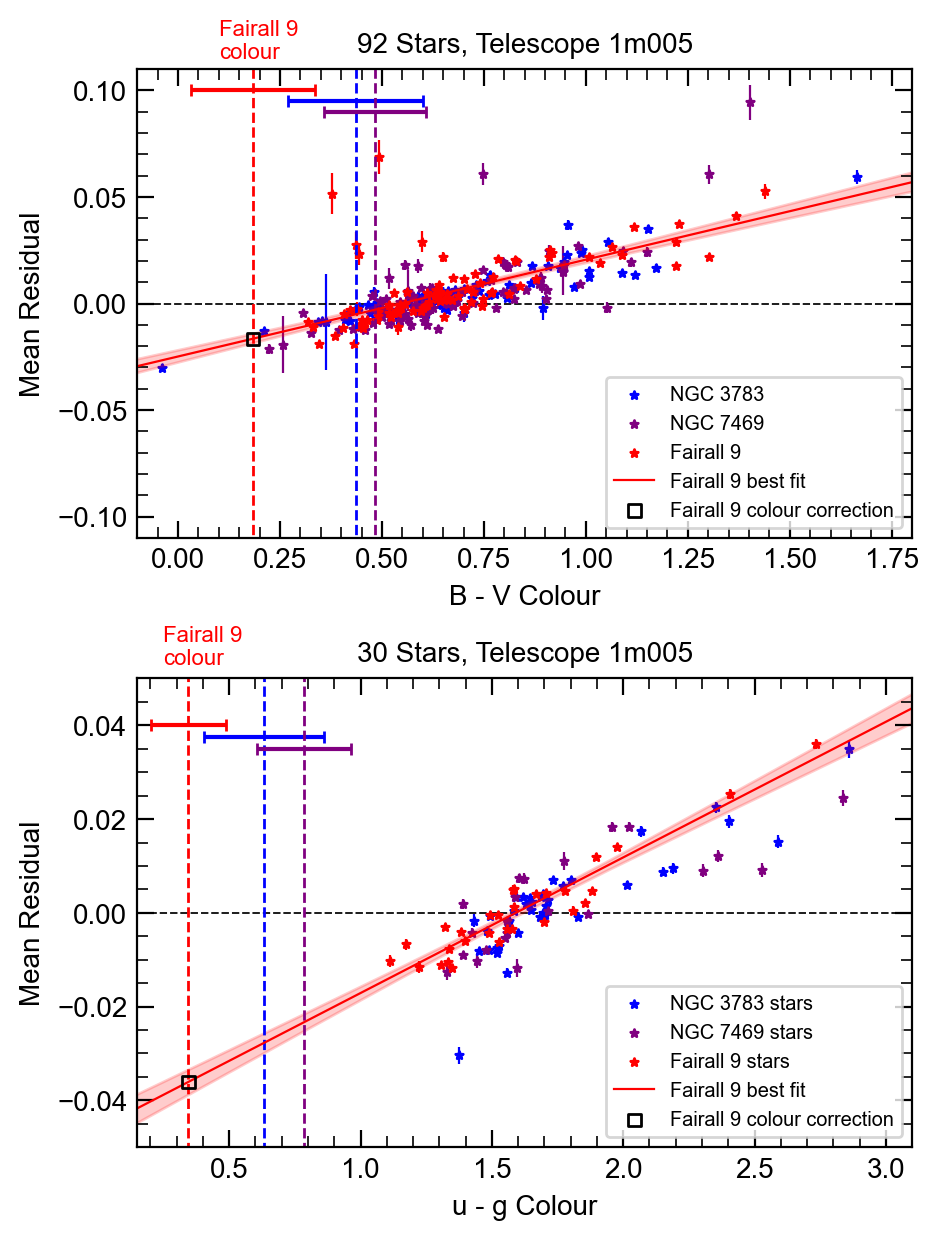} 
    \caption{
        \label{fig:colourcor}
        \textit{Top}: The mean (non-normalised) residuals as a function of the $B - V$ colour index for the field stars of Fairall\,9 (red), NGC\,3783 (blue), and NGC\,7469 (purple), in the LCO $B$ band. 
        Vertical dashed lines indicate the mean AGN colour, with horizontal errorbars showing the colour range throughout the observing period.
        The red solid line shows a linear fit to the Fairall\,9 data, extrapolated to the AGN colour where the value of the residual (black square) is used as a first-order colour correction for the telescope (1m005). 
        \textit{Bottom}: Same as above, but for the $u - g$ colour index. 
    Less stars are available across both bands for all three AGN (e.g, only 30 stars are available for Fairall\,9 as opposed to the 92 available across the $B$ and $V$ bands) and thus all the AGN colours appear blueward of their field stars.
    }
\end{figure}

\begin{figure}
    \centering
    \includegraphics[width=\columnwidth]{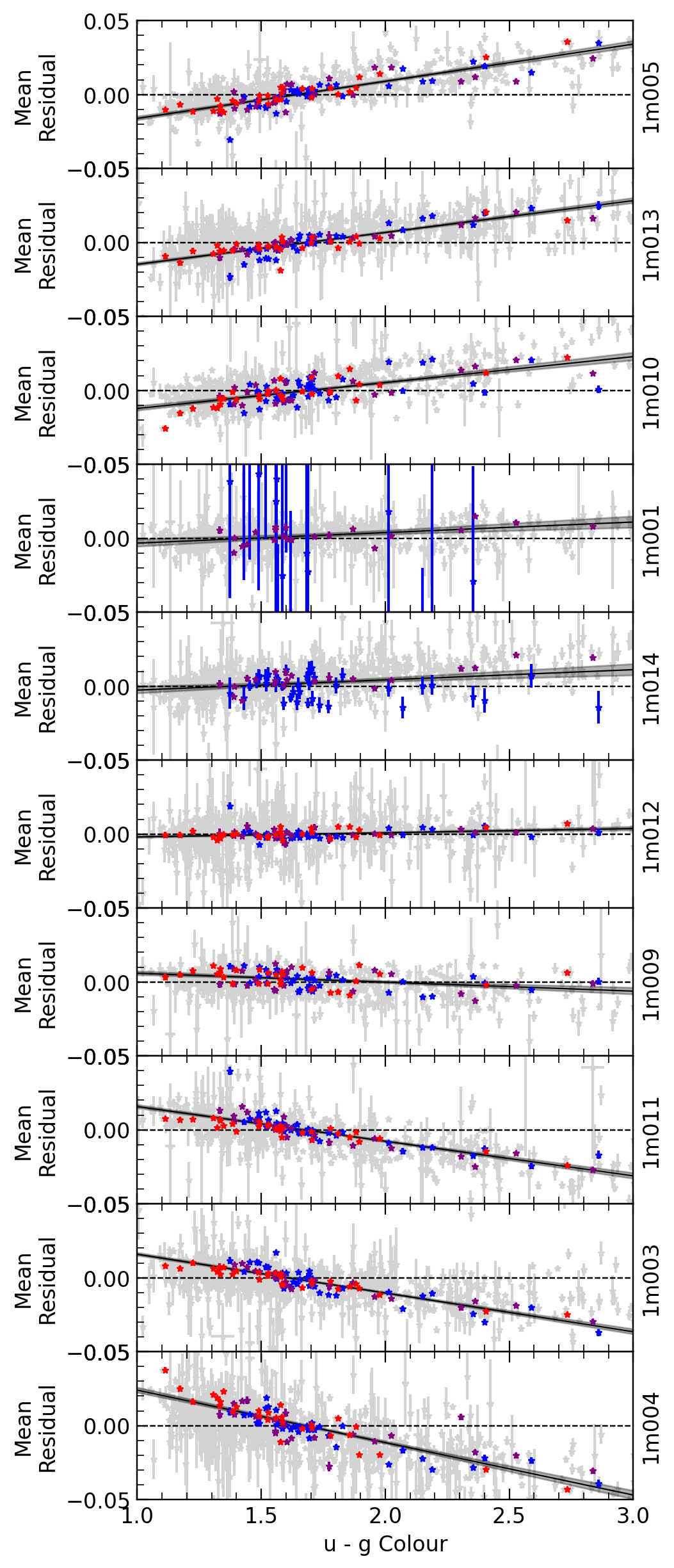} 
    \caption{\label{fig:allscopes} Same as Fig\,\ref{fig:colourcor} (bottom) for 12 LCO 1-m telescopes in the LCO $B$ band but with linear fits (black) performed with combined datapoints from NGC\,3783 (blue), NGC\,7469 (purple), and Fairall\,9 (red). The plots from top to bottom are ordered from the largest positive slope to the largest negative slope. Telescopes 1m001 and 1m014 generally have very few datapoints in each lightcurve, as seen for NGC\,3783 in Fig\,\ref{fig:heatmap}. In light gray we show the mean residuals for the other 29 AGNs in our LCO database.}
\end{figure}
In comparison, we show the same lightcurve section calibrated with {\tt PyTICS} (Fig.\,\hyperlink{Figpyroa}{\ref{fig:pyroa}} bottom). The variability patterns are comparable between the two methods but the uncertainty expansion is clearly different. The errorbars for telescope 1m009 are on average larger than from {\tt PyROA}, with no capping above an arbitrary threshold, and the outliers in the gaps are also sufficiently expanded since they are independently computed from the scatter in the comparison star data. {\tt PyROA} determines one extra variance parameter per telescope, meaning that a problematic telescope will tend to have a relatively larger telescope-specific variance. In this case, however, telescope 1m009 is only problematic at certain epochs due to the focus issues, and data outside this period is generally fine. {\tt PyTICS} is able to recognize this and adds a larger epoch-specific extra variance parameter $\sigma_{\rm Ep}$ in the noise model (Eq.\,\hyperlink{Eqvar}{\ref{eq:var}}) with no effect on $\sigma_{\rm Tel}$. {\tt PyROA} only effectively minimizes the impact of the bad frames if the scatter crosses the sigma-clipping threshold. In the {\tt PyROA} calibrated lightcurve, we therefore see the non-problematic 1m009 datapoints with error-bars that are larger than suggested by the short-timescale variability, as well as by the neighbouring data from other telescopes.

Fig.\,\hyperlink{Fighist}{\ref{fig:hist}}, examines the distribution of the total AGN noise model (Eq.\,\hyperlink{Eqvar}{\ref{eq:var}}) from both methods. {\tt PyTICS} has no regime in which the bad datapoints are fully dominated by the extra variance as seen in the sharp jumps in the {\tt PyROA} CDF. 
The median uncertainty for {\tt PyROA} is approximately 1.5 times smaller than {\tt PyTICS}, with 0.0039 and 0.0058 mag respectively, although {\tt PyROA} shows a bimodal distribution in the noise model. The uncertainty range for {\tt PyTICS} is 0.0015 to 0.39 mag, in contrast with {\tt PyROA} which has a narrower range of 0.0032 to 0.15 mag. Rogue data points with underestimated uncertainties, such as the ones from {\tt PyROA} in Fig.\,\ref{fig:pyroa}, have been seen to distort the cross-correlation function (CCF) used to obtain delay measurements in reverberation mapping. This does not only affect the lag measurements but the lag uncertainties as well, which can in turn result in inaccurate disc size measurements when fitting weighted parametric models to the lag spectra. If certain telescopes have inaccurately larger uncertainties in the lightcurve compared to the other telescopes (like 1m009), their contribution will be down-weighted when modelling the variability with more sophisticated delay measurement tools, distorting the variability properties of the AGN.

Moreover, although {\tt PyROA} makes no assumption about the underlying AGN variability (unlike {\tt PyCALI}), it still has a parametric model dependence with the ROA, which smooths the variability based on the width and shape of the running window function. This can flatten the shortest timescale peaks and troughs in the lightcurve, introducing biases at high frequencies which may lead to differences in power spectral density (PSD) analysis. Studies of the PSDs of quasars will require a model-independent calibration to extract the underlying PSD.
Additionally, large gaps in the data, like seasonal gaps, can introduce bias to the ROA model on the edges of the lightcurve, which again distorts the true variability pattern of the AGN. In contrast, {\tt PyTICS} computes the correction parameters independent of the AGN lightcurve, completely avoiding any structured modelling of the stochastic variability.

\hypertarget{Section4}{}
\section{Residual Trends with Star Colour}

A Gaussian distribution centred at zero for the normalised residuals indicates that the noise model adequately accounts for the scatter in the data (see right panel in Fig.\,\hyperlink{Figstarcal}{\ref{fig:starcal}}). 
In more detail, a heatmap of the normalised residuals for each star and epoch should resemble a white-noise pattern. 
Sorting the epochs by telescope and stars by colour, however, we find clear residual trends in the heatmap for each telescope (Fig.\,\hyperlink{Figheatmap}{\ref{fig:heatmap}}), 
which are overlooked if only the residual histogram is examined. The colours are computed as the difference in mean magnitude between two filters, after a zero-point correction is made to the calibrated instrumental magnitude lightcurves to bring them to a standard magnitude system.
We find no clear residual trends when sorting by other properties such as airmass or star magnitude, seeing instead the expected random pattern. 
Despite the clear colour-dependent trends, the {\tt PyTICS} intercalibrations of NGC\,3783 and several other AGN in our database appear to be not unduly compromised.

The evident colour-dependent residual trends are likely a consequence of the LCO 1-m telescopes having somewhat different wavelength-dependent sensitivities,
so that some are more responsive to relatively red and others to relatively blue comparison stars. 
Since we find no clear trend with the airmass, the colour-dependent effects are likely not due to atmospheric extinction which is also sensitive to wavelength. 
The epoch-specific corrections computed from our algorithm should calibrate out the main atmospheric extinction effects, at least for sources similar to the available comparison stars.
There will be differences in the extinction curves for each LCO site (e.g., different water vapour content in the observatories, different dust content etc.) but this would be hard to quantify without moving CCD cameras to the different sites. 

Considering as an example comparison stars in the fields of NGC\,3783, NGC\,7469, and Fairall\,9, their colour-dependent (non-normalised) residuals\footnote{The mean of the magnitude residuals across all epochs for a given telescope and star, as seen in the distinct rows of Fig.\,\hyperlink{Figheatmap}{\ref{fig:heatmap}}} can be examined in Fig\,\ref{fig:colourcor} for LCO telescope 1m005, and in Fig\,\ref{fig:allscopes} for all 12 LCO 1-m telescopes. For each telescope, the magnitude residuals as a function of star colour index have slopes that are similar for the comparison stars in different AGN fields, giving more evidence that this is likely a systematic effect due to differences in the spectral responses across telescopes. As shown in Table~\ref{tab:my_label}, telescope 1m004 and 1m005 both originate at the same site, the Cerro Tololo Interamerican Observatory in Chile, and show colour-residual trends with opposite slopes.

Fernández Fernández et\,al. (\citeyear{fernandez_fernandez_improved_2012}) compares several methods that build upon conventional differential photometry, which typically selects a few comparison stars in the same spectral range as the target to minimise the wavelength-dependent atmospheric extinction. 
Their new method models the lightcurves of auxiliary stars, that are not necessarily in the same spectral range as the target, with a polynomial to remove the strong trends with elevation (changes with airmass) before constructing a comparison star lightcurve to calibrate the target. 
Our observations are of approximately daily cadence where the airmass changes more randomly and as a result such smooth trends are not present in the data. {\tt PyTICS} computes and removes the epoch-specific offsets while considering the scatter across the comparison stars, and therefore takes into account the small differences across the star spectral types. Furthermore, in Fernández Fernández (\citeyear{fernandez_fernandez_improved_2012}), they compare their results to the methods of Tamuz et al. (\citeyear{tamuz_correcting_2005}), which explicitly uses the airmass term in the calibration solution. 
As stated in their study, however, the correction parameters likely have no relation to the actual airmass values.

\begin{figure}
    \centering
    \includegraphics[width=\columnwidth]{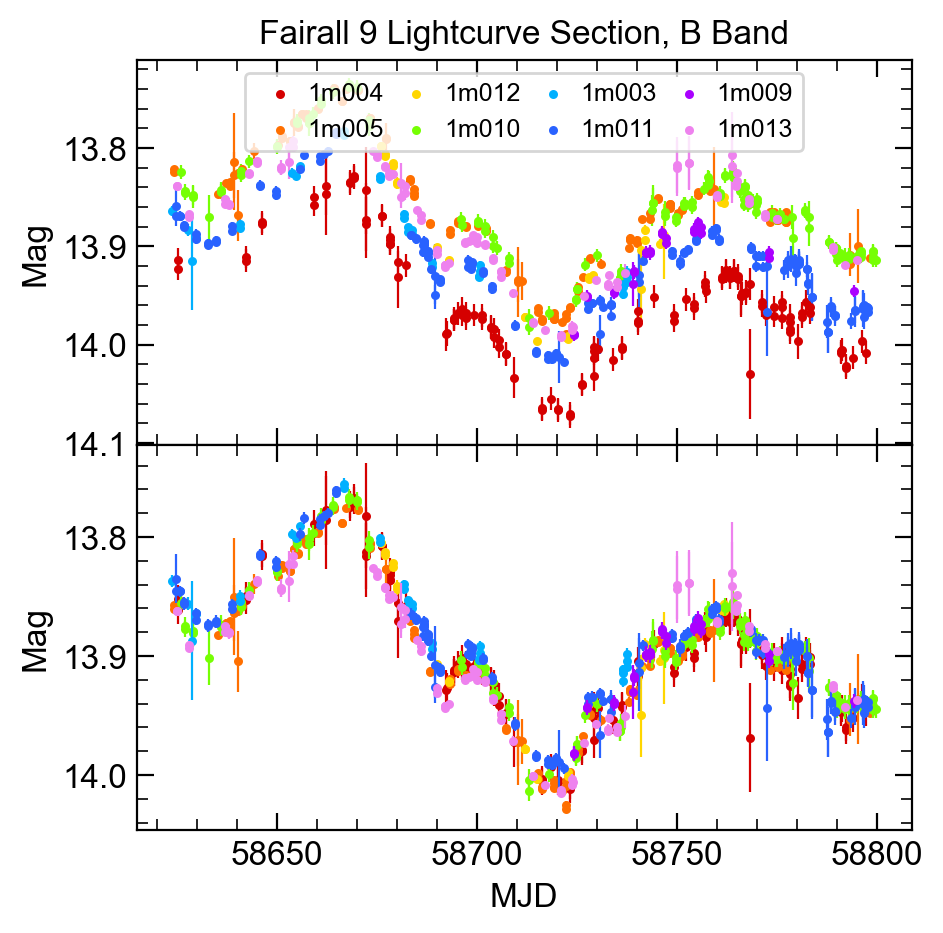} 
    \caption{
    \label{fig:f9}
    \textit{Top:} A section of the Fairall\,9 lightcurve in the LCO $B$ band, coloured by telescope, showing strong splitting even after using {\tt PyTICS}. 
    A first-order colour correction can be derived from fitting the mean (non-normalised) residuals as a function of star colour and extrapolating to the AGN colour. \textit{Bottom}: The same lightcurve section after applying the first-order colour correction parameters, using the $u-g$ colour index.}
\end{figure}
\begin{figure}
    \centering
    \includegraphics[width=\columnwidth]{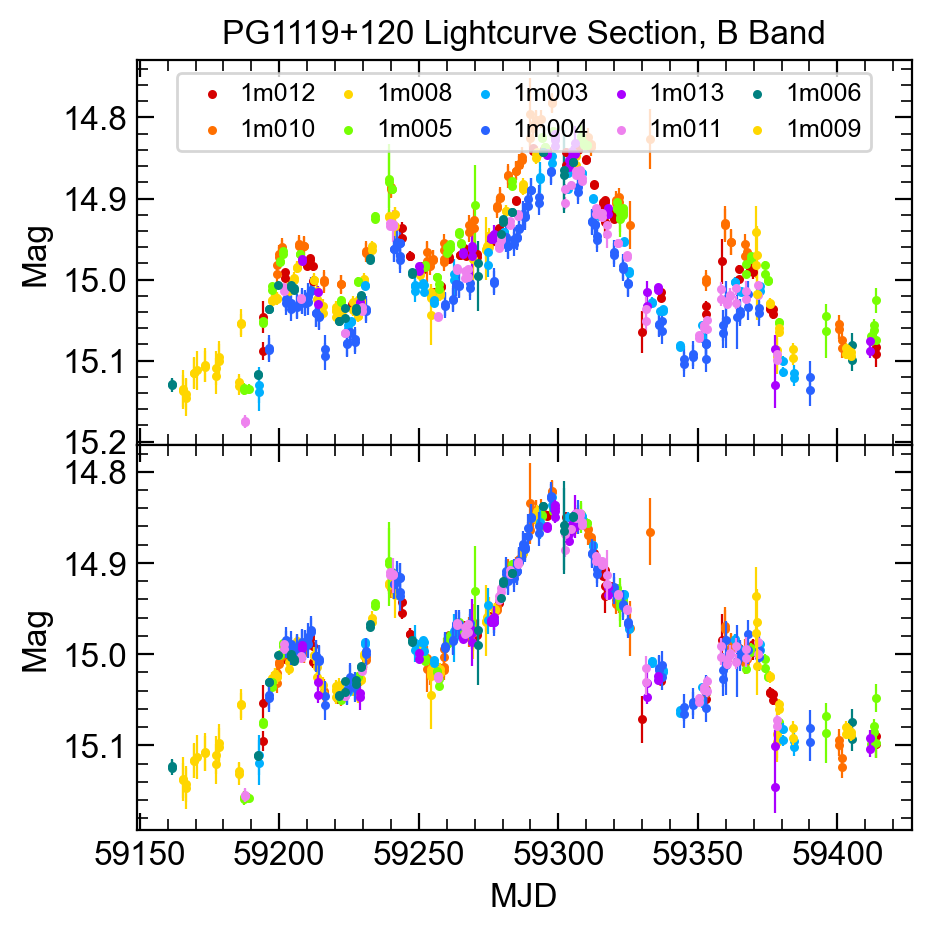} 
    \caption{Same as Fig \ref{fig:f9} but for PG\,1119+120.}
    \label{fig:PG}
\end{figure}
\subsection{Colour Terms for very Blue AGN}

A demonstrative case where the colour-dependent trends are important is the relatively blue AGN Fairall\,9, which seems to struggle with the {\tt PyTICS} calibration. 
In the top panel of Fig.\,\hyperlink{Figf9}{\ref{fig:f9}}, we show a section of the Fairall\,9 lightcurve in the $B$ band after using {\tt PyTICS}, where the `splitting' effect in the lightcurve remains strong. 
Examining the distribution of star colours in the field of Fairall\,9, we find $B - V$ values ranging between 0.30 and 1.45 for the 92 stars that are available in both bands, whereas Fairall\,9 has a mean $B - V$ colour index of 0.18, sitting blueward of all the calibration stars (Fig.\,\hyperlink{Figcolourcor}{\ref{fig:colourcor}} top). 
For NGC\,3783 for example, the mean $B - V$ colour index is 0.44, sitting within its field star colour range of 0.25 to 1.25.  We note that for the other colour indices, fewer comparison stars are available in both bands, and thus the AGN colours all appear blueward of their comparison stars (Fig.\,\hyperlink{Figcolourcor}{\ref{fig:colourcor}} bottom; e.g., only 30 stars are available for the $u - g$ colour index), but Fairall\,9 remains significantly bluer.

To further investigate this difference in calibration quality between NGC\,3783 and Fairall\,9, we have used {\tt PyTICS} on 32 AGNs with available \textit{B} band data in our LCO database. We find three additional objects\footnote{PG 1119+120, Mrk\,1044, and 3C\,273} that exhibit a telescope-specific splitting effect and show the same trend as Fairall\,9; they all have a relatively bluer $B-V$ colour index compared to their comparison stars. The remaining 84\% of the AGNs sit within the colour index range of their field stars and were successfully calibrated using {\tt PyTICS}, with comparable quality to NGC\,3783.

As a first-order colour correction, we perform linear fits to the mean (non-normalised) telescope residuals as a function of star colour (Fig.\,\hyperlink{Figcolourcor}{\ref{fig:colourcor}}) using {\tt linmix}\footnote{\url{https://linmix.readthedocs.io/en/latest/src/linmix.html}} (Kelly \citeyear{Kelly2007}), which implements Bayesian linear regression, accounting for uncertainties in both the star colour and the magnitude residuals. 
We then extrapolate the linear model fit to the colour of the AGN and use this residual value as an additional colour correction parameter $\Delta m_{\text{Col, Tel}}$. 
Computing this correction for each telescope and applying it to the Fairall\,9 data reduces the telescope-specific offsets significantly, as shown in the bottom panel of Fig.\,\hyperlink{Figf9}{\ref{fig:f9}}. 
The largest colour correction parameter for Fairall\,9 in the \textit{B} band is that of telescope 1m004 with $\Delta m_{\text{Col, Tel}}$ = 0.061 $\pm$ 0.004, compared to NGC\,3783 and NGC\,7469 with 
$\Delta m_{\text{Col, Tel}}=0.036\pm0.003$ 
and $\Delta m_{\text{Col, Tel}}=0.020\pm0.006$, respectively. This colour-correction method also successfully reduces the systematic offsets for the three additional AGNs that show significant splitting (see second example for PG\,1119+120 in Fig.\,\ref{fig:PG}).

Given that the slopes appear consistent across field stars of different AGNs (see Fig.\,\hyperlink{Figallscopes}{\ref{fig:allscopes}}), the mean residuals as a function of star colour can be combined from all the datasets to derive analytical models of the first-order correction parameters as a function of AGN colour for the entire LCO telescope network. In Appendix \ref{sec:A} we compare the colour correction results for the four blue AGNs when using this global solution versus when using only the specific field stars of each AGN. We find that the solution derived from our 32 AGN fields corrects the lightcurves to a similar degree as the individual fields (see Fig.\,\ref{fig:32AGN}), but with consistent deviations for telescopes 1m004 and 1m010. Ultimately, correcting individual AGNs using their own field stars yields better quality colour corrections compared to the global solution.


A caveat for the colour correction method is the dependency of the colour index used to derive the colour correction parameters.
The colour correction result shown in Fig.\,\hyperlink{Figf9}{\ref{fig:f9}} is performed using the $u - g$ colour index, and although the residual trends are seen across other colour indices that we considered, they did not perform as well in correcting the lightcurves of the bluer AGNs. 
The choice of colour index is thus somewhat arbitrary and may not work as well if only certain filters are available. The colour-dependent problem occurs for only a small fraction of our AGNs, and the first-order colour correction method is able to mitigate the systematic offsets when using the $u-g$ colour index. If the `splitting' effect of bluer AGNs cannot be sufficiently corrected in this way, we recommend combining the intercalibration with {\tt PyTICS} and {\tt PyROA}; the former to compute a robust noise model and the latter to reduce the telescope-specific offsets in the data. Alternatively, to optimise the additional colour-dependent correction parameter, including a colour-dependent extra variance parameter in the AGN noise model, this process could be incorporated into the iterative algorithm. 
This would involve performing Steps\,1-3 for at least two filters simultaneously; ones that are used to determine the colour index. We leave this implementation for future work.

Lastly, unlike the comparison stars, AGN exhibit a `bluer when brighter' behaviour, meaning that their colour indices change throughout the observing period. In Fig.\,\ref{fig:colourcor}, horizontal error bars on the mean AGN colours show the colour range for the entire lightcurve, which consequently gives a range of colour correction parameters.
We note here that the variability amplitude of the AGN is smaller for the $u - g$ colour index than for the other colour indices that we considered, perhaps because contamination of the relatively blue AGN by relatively red host galaxy starlight is minimised in the $u$ band.
The AGN magnitude as a function of colour therefore shows a tight linear relation that we fit to derive a model for the AGN colour index as a function of time (Fig.\,\ref{fig:bluerbrighter})
Using this time-dependent AGN colour to obtain the correction parameters from the linear model of the mean residuals, we a difference of <0.002 in magnitudes between lightcurves calibrated using the mean AGN colour and lightcurves calibrated using the time-dependent colour (see Fig.\,\ref{fig:magresl}). The `bluer when brighter' nature of AGN is thus not a significant factor in the quality of the colour correction, which is most sensitive to the residual-colour slopes.
\hypertarget{Section5}{}
\section{Summary}

We present an iterative optimal scaling algorithm for intercalibrating AGN lightcurves that combine data from different telescopes, using 100s of comparison stars on the same images. 
The algorithm uses maximum likelihood estimation to derive epoch-specific and telescope-specific correction parameters, along with the corresponding extra variance parameters. 
The AGN lightcurve is then corrected at the end of the iterative process, requiring no interpolation, and thus no assumptions about AGN variability. 
Applied to data from the Las Cumbres Observatory (LCO) 1-m robotic telescope network, 
the resulting intercalibrated AGN lightcurves are comparable to those found by using other intercalibration methods.
The uncertainty estimation based on the scatter in the comparison star data, and in particular the identification and treatment of outliers, is improved compared with methods that intercalibrate using the AGN lightcurve. 
Our iterative optimal scaling algorithm also requires far less compute time than methods that use MCMC to sample the full joint distribution of the calibration parameters.

We find residual trends when sorting the star data by colour and the epochs by telescope, which are likely due to wavelength-dependent sensitivity differences across the LCO 1-m telescopes. 
These colour-dependent trends do not significantly compromise the quality of the calibration for NGC\,3783, but are important for Fairall\,9.
A first-order colour correction can reduce the telescope-specific shifts for AGN that are significantly bluer than their comparison stars, as we have demonstrated for Fairall\,9.

The algorithm is available as a public-access Python module {\tt PyTICS}\footnote{\url{https://github.com/Astroberta/PyTICS}}. 
This allows for straightforward intercalibration of time-series observations using the methods outlined here. 
The package also contains several diagnostic tools to examine the data post-calibration, including the residual heatmaps and RMS versus magnitude plots to identify particularly variable stars, which one may wish to omit if the number of stars is limited. Furthermore, the first-order additional colour corrections can be made, given data in at least two photometric bands.

\section*{Acknowledgements}
The authors thank the anonymous referees for their constructive comments, which helped improve the quality of the paper. This work makes use of observations from the Las Cumbres Observatory global telescope network.
RV acknowledges support from STFC studentship ST/Y509589/1. JVHS acknowledges support from STFC grant ST/V000861/1. 
This research made extensive use of {\sc matplotlib} (Hunter \citeyear{Hunter:2007aa}) and {\sc pandas} \citep{Pandas2022}.


\section*{Data Availability}
The raw datasets were derived from sources in the public domain at LCO archive \url{https://archive.lco.global}. The calibrated lightcurves are available from the author upon reasonable request.


\bibliographystyle{mnras}
\bibliography{References_abrv}

\appendix
\section{{\tt PyTICS} Analysis of 32 AGN Fields} We have used {\tt PyTICS} to calibrate all 32 AGNs with available $B$ band data in our LCO database. Four out of the 32 objects exhibit significant telescope-specific offsets, compared to the rest of the sample (see Fig.\,\ref{fig:f9} and Fig.\,\ref{fig:PG}). These AGN are Fairall\,9, PG\,1119+120, 3C\,273, and Mrk\,1044. The remaining AGN were successfully calibrated with {\tt PyTICS}, showing comparable quality to NGC\,3783. The common denominator among these four objects is that they are all relatively bluer in the $B-V$ colour index compared to their comparison stars. The first-order colour correction method described in Section\,\hyperlink{Section4}{4} reduces the systematic splitting substantially for all four objects.

\begin{figure*}
    \centering
    \includegraphics[width=\linewidth]{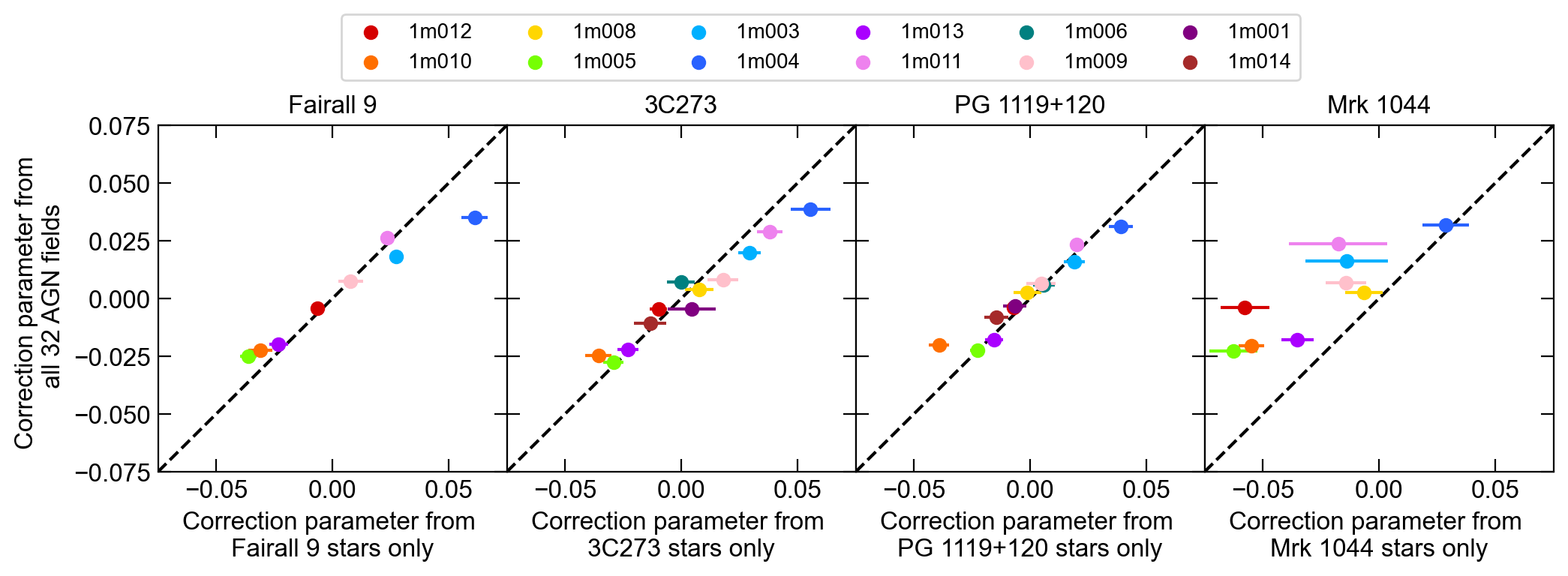}
    \caption{Comparison between colour correction parameters derived from the comparison stars of all 32 AGN fields and the field stars of the individual AGNs only, using the $u-g$ colour index, for the four relatively bluer AGNs in our LCO database. The points are coloured by telescope, with a one-to-one line shown in black.}
    \label{fig:32AGN}
\end{figure*}

Given that the slopes of the mean (non-normalised) residuals as a function of star colour appear consistent across the different AGN fields (Fig.\,\ref{fig:allscopes}), we compare in Fig.\,\ref{fig:32AGN} the colour-correction parameters derived from all 32 AGN comparison stars against the parameters derived from individual AGN field stars only. The parameters for Fairall\,9, 3C\,273, and PG\,1119+120 appear consistent with one another up to a $\sim$$2\sigma$ level, with some consistently larger deviations for telescopes 1m004 and 1m010. Mrk\,1044 shows less consistency between the parameters. A visual inspection of the colour-corrected lightcurve for Mrk\,1044 indicates that using the parameters derived from individual AGN field stars yields a better result than the global solution. The global solution is still a useful tool to have, requiring only the AGN colour to perform the colour correction, and we include it for all LCO telescopes and filters in our database alongside the {\tt PyTICS} python library at \url{https://github.com/Astroberta/PyTICS}.

\label{sec:A}

\section{Time Dependent AGN Colour}
\begin{figure}
    \centering
    \includegraphics[width=\linewidth]{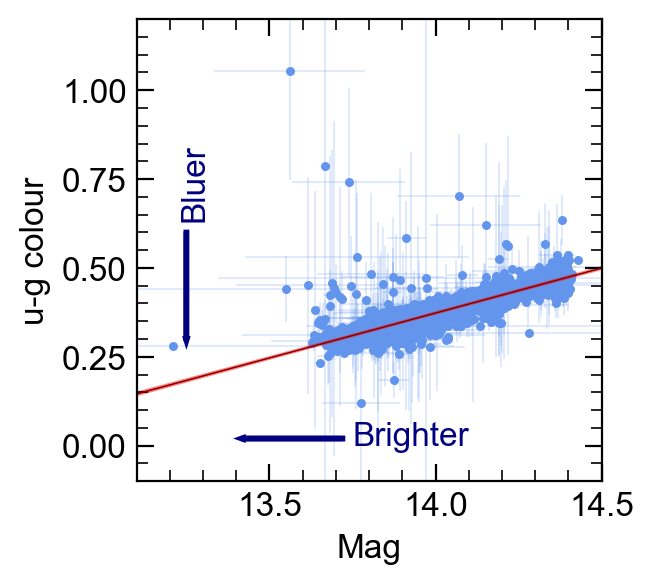}
    \caption{The $u-g$ colour index of Fairall\,9 plotted against the $B$ band magnitude for each epoch, highlighting the `bluer when brighter' behaviour of the AGN. A time-dependent AGN colour is derived from a linear fit to the data (red).}
    \label{fig:bluerbrighter}
\end{figure}

\begin{figure}
    \centering
    \includegraphics[width=\linewidth]{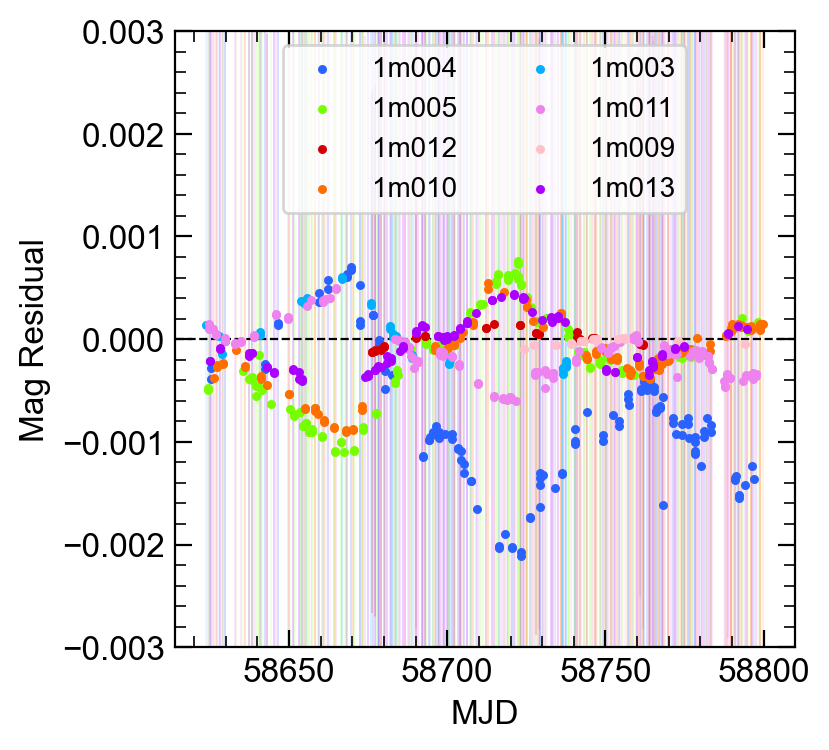}
    \caption{Magnitude residuals between the colour-corrected Fairall\,9 lightcurve segment when using the mean AGN colour and when using the time-dependent AGN colour, showing a difference between the two lightcurves of less than 0.002 mag.}
    \label{fig:magresl}
\end{figure}

The first-order colour correction method described in Section\,\hyperlink{Section4}{4} requires the colour index of the object. AGN exhibit a `bluer when brighter' behaviour, and thus the correction parameter for each telescope will vary depending on the colour of the AGN at a given epoch. We investigate the significance of this effect by deriving time-dependent colour correction parameters for Fairall\,9. This is achieved by performing a linear fit to the AGN colour index as a function of magnitude at each epoch, as shown in Fig.\,\ref{fig:bluerbrighter}. Using this magnitude-dependent (and thus time-dependent) AGN colour, the correction parameters are derived in the same way from the residual-colour trends. As shown in Fig.\,\ref{fig:magresl}, the difference in magnitude between the lightcurve corrected using the mean AGN colour and the time-dependent colour is less than 0.002. From our sample of blue AGNs we can therefore conclude that the `bluer when brighter' nature is not a significant factor in the colour correction, however this option should not be dismissed if more bluer AGNs are found with this issue. The {\tt PyTICS} library has an option to compute these time-dependent colour correction parameters if desired.

\label{sec:B}


\bsp	
\label{lastpage}
\end{document}